\documentclass[12pt]{article}

\def\lsim{\mathrel{\rlap {\raise.5ex\hbox{$ < $}}
{\lower.5ex\hbox{$\sim$}}}}
\def\gsim{\mathrel{\rlap {\raise.5ex\hbox{$ > $}}
{\lower.5ex\hbox{$\sim$}}}}
\begin{document}

\begin{titlepage}

\begin{centering}
\vspace{.6in}
{\LARGE  \bf {New Minimal Extension of MSSM} }\\
\vspace{2 cm}
{\large C. Panagiotakopoulos$^a$
 and K. Tamvakis$^b$}\\
\vskip 1cm
{$^a$\it{Theory Division, CERN, CH-1211 Geneva 23, Switzerland\\
and\\
Physics Division, School of Technology\\
Aristotle University of Thessaloniki, 54006 Thessaloniki, Greece}\\
\vskip 1cm
{$^b$\it {Physics Department, University of Ioannina\\
45110 Ioannina, Greece}}}\\

\vspace{1.5cm}
{\bf Abstract}\\
\end{centering}
\vspace{.1in}
We construct a new   minimal extension of the Minimal   Supersymmetric
Standard Model (MSSM) by  promoting  the $\mu$-parameter to  a singlet
superfield.  The  resulting renormalizable  superpotential is enforced
by   a   $\mathcal{Z}_5$   $R$-symmetry  which  is     imposed  on the
non-renormalizable operators  as well.  The  proposed model provides a
natural   solution   to   the   $\mu$-problem   and   is  free    from
phenomenological and cosmological problems.

\vspace{1cm}
\begin{flushleft} 
August 1999
\end{flushleft}
\hrule width 6.7cm \vskip.1mm{\small \small}
 \end{titlepage}

The Minimal Supersymmetric extension
of the Standard Model (MSSM) \cite{NHK}, defined by promoting each standard 
field to a superfield, doubling the higgs fields and imposing $R$-parity
conservation, seems to be preferred by the low energy data
which support unification of the gauge couplings in the supersymmetric case.
The most viable scenario for the breaking of supersymmetry at
some low scale $m_{s}$, no larger than $\sim $ $1\ TeV,$ is the one based on
spontaneously broken supergravity. The breaking of supergravity takes place 
in some hidden sector and is communicated to the visible sector through 
gravitational interactions. The resulting theory with broken supersymmetry 
contains,
independently of the details of the underlying high energy theory,
a number of \textit{soft} supersymmetry (susy) breaking terms proportional
to powers of the scale $m_{s}$. Probably the most attractive feature of the
MSSM is that it realizes a version of ``dimensional transmutation'' where
radiative corrections generate the electroweak scale $M_{W}$ from the 
susy-breaking scale $m_{s}$. Unfortunately, a realistic implementation of 
radiative symmetry breaking \cite{RSB} in MSSM requires the presence of a 
coupling $\mu H_{1}H_{2}$ involving the higgs fields $H_{1}$ and $H_{2}$, 
the so called $\mu$ term, with values of the theoretically arbitrary parameter 
$\mu $ close to $m_{s} \sim M_{W}$.
This nullifies all merits of radiative symmetry breaking since it amounts to
introducing the electroweak scale by hand.  
Of course, there exist scenarios to account for 
the origin of the $\mu$ term, alas, all in extended settings \cite{MU}.

A straightforward solution to the $\mu $-problem would be to enlarge the field
content of MSSM by adding
a massless gauge singlet field $S$ that couples to the higgs fields 
as $\lambda SH_{1}H_{2}$ and acquires a vacuum expectation value (vev) 
of the order of $m_{s} \sim M_{W}$.
Such a  model with a purely cubic renormalizable superpotential 
containing a self-coupling of $S$ as well became known as the 
``Next to Minimal'' SSM or NMSSM \cite{NM}. 
At the renormalizable level the model possesses a ${\mathcal{Z}}_{3}$ symmetry 
under which all 
superfields are multiplied by $e^{2\pi i/3}$ whose spontaneous breaking
leads to the formation of cosmologically catastrophic domain walls unless
the discrete symmetry is not respected by higher order (non-renormalizable)
operators. The existence of higher order operators
\footnote{%
These non-renormalizable terms appear either as $D$-terms in the K{\"{a}}%
hler potential or as $F$-terms in the superpotential. The natural setting
for these interactions is $N=1$ Supergravity spontaneously broken by a set
of hidden sector fields.}
violating the 
${\mathcal{Z}}_{3}$ symmetry, however, was shown \cite{ASW} to be intimately
related to the generation of quadratically divergent tadpoles for the 
singlet \cite{LAX}. 
Their generic contribution to the effective potential, cut-off at the Planck
scale $M_{P}$, is
\begin{equation}
\delta V \sim \xi M_{P}m_{s}^{2}S + h.c.,
\end{equation}
where $\xi$ is a factor depending on the loop order in which the tadpole
is generated. Such terms tend to destabilize the gauge hierarchy
through a vev for the light singlet $S$ much larger than the electroweak scale.

Recently we have found \cite{PT} a simple resolution to the above problems of
NMSSM by imposing a ${\mathcal{Z}}_{2}$ R-symmetry on the 
non-renormalizable operators under which all superfields as well as the
superpotential flip sign. Thus, the potentially harmful to the gauge 
hierarchy operators \cite{ABEL} are forbidden but a harmless tadpole  
\begin{equation}
\delta V \sim \xi m_{s}^{3}S + h.c.
\end{equation} 
breaking the ${\mathcal{Z}}_{3}$ symmetry and making the walls disappear
can still be generated.  

Our purpose in the present note is to get rid of the cubic superpotential
self-coupling of the singlet $S$ thereby constructing the simplest extension
of the MSSM. To accomplish our goal we should, of course, find substitutes
for the twofold role played by the $S^{3}$ coupling, as this trilinear coupling
contributes to the mechanism generating the vev of $S$ through the soft 
susy-breaking terms and explicitly breaks the unwanted Peccei-Quinn symmetry 
present in its absence.

The renormalizable superpotential of the proposed model   
\begin{equation}
{\mathcal{W}}_{ren}={\lambda }SH_{1}H_{2}+%
Y^{(u)}QU^{c}H_{1}+Y^{(d)}QD^{c}H_{2}+Y^{(e)}LE^{c}H_{2}
\end{equation}
possesses the global symmetries
\[
U(1)_{B}:Q(\frac{1}{3}),\,U^{c}(-\frac{1}{3}),\,D^{c}(-\frac{1}{3}%
),\,L(0),\,E^{c}(0),\,H_{1}(0),\,H_{2}(0),\,S(0) ; 
\]
\[
U(1)_{L}:\,Q(0),\,U^{c}(0),\,D^{c}(0),\,L(1),\,E^{c}(-1),\,H_{1}(0),%
\,H_{2}(0),\,S(0) ;
\]
\[
U(1)_{PQ}:\,Q(-1),\,U^{c}(0),\,D^{c}(0),\,L(-1),\,E^{c}(0),\,H_{1}(1),%
\,H_{2}(1),\,S(-2) ;
\]
\[
U(1)_{R}:\,Q(1),\,U^{c}(1),\,D^{c}(1),\,L(1),\,E^{c}(1),\,H_{1}(0),%
\,H_{2}(0),\,S(2), 
\]
where the charge of the superfield under the corresponding symmetry is given 
in parenthesis. 
$U(1)_{B}$ and $U(1)_{L}$ are the usual baryon and lepton number symmetries,
$U(1)_{PQ}$ is an anomalous Peccei-Quinn symmetry whereas $U(1)_{R}$
is a non-anomalous $R$-symmetry under
which the renormalizable superpotential ${\mathcal{W}}_{ren}$ has charge 2.
The soft trilinear susy-breaking terms break the continuous $R$-symmetry $%
U(1)_{R}$ down to its maximal non-$R$ ${\mathcal{Z}}_{2}$ subgroup 
which is the usual matter-parity.
The $U(1)_{PQ}$, which remains unbroken by the soft susy-breaking terms,
could be broken by a linear effective potential term of the type
given by Eq. (2) with $\xi \sim 1$ arising from non-divergent tadpoles.
It is, however, quite difficult to achieve such an unsuppressed value of $\xi$.
Thus, we rather have to resort to divergent tadpole contributions cut-off at
$M_{P}$ which occur at very high order such that $\xi M_{P} \sim m_{s}$. 
Finally, $U(1)_{B}$ and $U(1)_{L}$ remain unbroken by both the susy-breaking
terms and the tadpole but might be violated by some non-renormalizable 
operators hopefully of sufficiently high order. Consequently, it is
sufficient to find a symmetry which ensures the renormalizable superpotential
of Eq. (3) and allows the generation of an adequately suppressed tadpole.
All unwanted symmetries will then be broken and a vev $<S> \sim m_{s}$ will
readily be generated by combining the soft susy-breaking mass-squared term 
$\sim m_{s}^{2}SS^{*}$ with the above linear in $S$ contribution to the 
effective potential.

A continuous symmetry enforcing the form of ${\mathcal{W}}_{ren}$ in   
Eq. (3) is the $U(1)$ $R$-symmetry obtained as the linear combination 
$R'=3R+PQ$ of $U(1)_{R}$ and $U(1)_{PQ}$ :   
\[
U(1)_{R'}:\,Q(2),\,U^{c}(3),\,D^{c}(3),\,L(2),\,E^{c}(3),\,H_{1}(1),%
\,H_{2}(1),\,S(4) 
\]
under which the superpotential ${\mathcal{W}}$ has charge 6. 
$U(1)_{R'}$ is broken by the
trilinear soft susy-breaking terms down to its maximal non-$R$ subgroup
${\mathcal{Z}}_{6}$ which is the product of a ${\mathcal{Z}}_{2}$ and a 
${\mathcal{Z}}_{3}$ subgroup. The ${\mathcal{Z}}_{2}$
is essentially the usual matter parity (up to a $SU(2)_{L}$ element reversing
the sign of all doublets) which leaves the tadpole invariant. Under the 
${\mathcal{Z}}_{3}$ (which is a subgroup of $U(1)_{PQ}$) instead, $S$ 
transforms non-trivially and the tadpole does not remain invariant. 
Thus, we should avoid imposing the whole $U(1)_{R'}$ symmetry or one 
of its subgroups which contains the aforementioned ${\mathcal{Z}}_{3}$ if we 
want a tadpole to be generated.

A subgroup of $U(1)_{R'}$ which is completely broken by the trilinear soft
susy-breaking terms and is sufficiently large to enforce 
${\mathcal{W}}_{ren}$ of Eq. (3) but sufficiently small to allow the generation
of a sizeable tadpole is the ${\mathcal{Z}}_{5}$ subgroup 
${\mathcal{Z}}_{5}^{r}$ of $U(1)_{R'}$ generated by
\[
{\mathcal{Z}}_{5}^{r}\,:\,(H_{1},H_{2})\rightarrow \alpha(H_{1},H_{2}),\ \ \,
(Q,L)\rightarrow {\alpha}^{2}(Q,L),\ \ \,
(U^{c},D^{c},E^{c})\rightarrow
{\alpha}^{3}(U^{c},D^{c},E^{c}),\ \ \,
\]
\[
S\rightarrow {\alpha}^{4}S,\ \ \,\mathcal{W}\rightarrow \alpha \mathcal{W}
\]
with $\alpha = e^{2 \pi i/5}$. To examine the generation of the tadpole we
bear in mind that the potentially harmful non-renormalizable terms are either
even superpotential terms or odd K{\"{a}}hler potential ones. 
Moreover, a tadpole diagram
is divergent if an odd number of such ``dangerous'' non-renormalizable terms
is combined with any number of renormalizable ones. 
Respecting the above rules \cite{ABEL} and the 
${\mathcal{Z}}_{5}^{r}$ $R$-symmetry we were able to show, not without some 
effort, that divergent tadpoles first appear at six loops.
One example of such a divergent six-loop tadpole diagram is obtained by
combining the non-renormalizable K{\"{a}}hler potential terms 
${\lambda_{1}}S^{2}H_{1}H_{2}/M_{P}^{2} + h.c.$ and
${\lambda_{2}}S(H_{1}H_{2})^{3}/M_{P}^{5} + h.c.$ with the renormalizable
superpotential term ${\lambda}SH_{1}H_{2}$ (four times).
The so generated linear effective potential term  
\[
\delta V\sim (16\pi ^{2})^{-6}{\lambda_{1}}{\lambda_{2}}{\lambda^{4}}%
M_{P}m_{s}^{2}S+ h.c. 
\]
is of the desired order of magnitude.

Notice that the ${\mathcal{Z}}_{5}$ $R$-symmetry 
${\mathcal{Z}}_{5}^{r}$, although it does not contain the usual matter parity, 
still manages to adequately stabilize the proton since, in addition to all 
$d=4$ baryon and lepton number violating operators, it also forbids the 
dangerous $QQQL$ and $U^{c}U^{c}D^{c}E^{c}$ $d=5$ ones.

It is very interesting that  non-zero light neutrino masses are
readily incorporated in the model by simply  introducing gauge singlet states
$\nu^{c}$ transforming  like $E^{c}$ under  all global symmetries. The
allowed large majorana mass terms for these  states break $U(1)_{L}$
down to its ${\mathcal{Z}}_{2}$  subgroup and generate small ordinary
neutrino masses through the standard see-saw mechanism.

In conclusion, we have shown that the $\mu$  term of  MSSM can be generated
by promoting  the   parameter $\mu$  to a
singlet superfield and   imposing a ${\mathcal{Z}}_{5}$  $R$-symmetry. 
The resulting model is a truly minimal extension of MSSM.

\subsection*{Acknowledgements}

We acknowledge support by the TMR network ``Beyond the Standard Model''. 
We also wish to thank A. Pilaftsis for his valuable comments
on the manuscript. C.P. wishes to thank S. Abel for many useful discussions.

\end{document}